# THE POWER SPECTRUM OF GALAXY DENSITY FLUCTUATIONS: CURRENT RESULTS AND IMPROVED TECHNIQUES[†]

Michael S. Vogeley
*Department of Physics and Astronomy, Johns Hopkins University*
*Baltimore, MD 21218 USA*
vogeley@pha.jhu.edu

**Abstract**

The power spectrum of density fluctuations measured from galaxy redshift surveys provides important constraints on models for the formation of large-scale structure. I review current results for the 3-D power spectrum and examine the limitations of current measurements and estimation techniques. To span the decade of wavelength between the scales probed by galaxy surveys and COBE, measure the detailed shape of the power spectrum, and accurately examine the dependence of clustering on galaxy species, we require deeper samples with carefully controlled selection criteria and improved techniques for power spectrum estimation. I describe a new method for estimating the power spectrum that optimally treats survey data with arbitrary geometry and sampling.

## 1 Introduction

Recent measurements of galaxy clustering from redshift surveys and angular catalogs, together with limits on the clustering of mass implied by the COBE DMR experiment, yield important constraints on proposed models for the formation of large-scale structure. However, we lack accurate constraints on fluctuations in galaxy density on scales that overlap with those probed by COBE, and the extant measurements have poor resolution on scales where certain theories predict interesting features in the power spectrum. Several surveys, either planned or in progress, promise to yield the desired measurements of the power spectrum of galaxy density fluctuations, but the complex geometry and sampling of these surveys pose a strong challenge to traditional methods of power spectrum analysis. The ultimate measurement of the galaxy fluctuation spectrum will result from combining all of the available data into one sample. This possibility begs the question, how do we obtain the best possible estimate of the power spectrum from a sample with arbitrarily complex geometry and sampling density?

---

[†]To appear in *Clustering in the Universe*, Proc. of XXXth Moriond Conference, Les Arcs, France, March 11-18, 1995

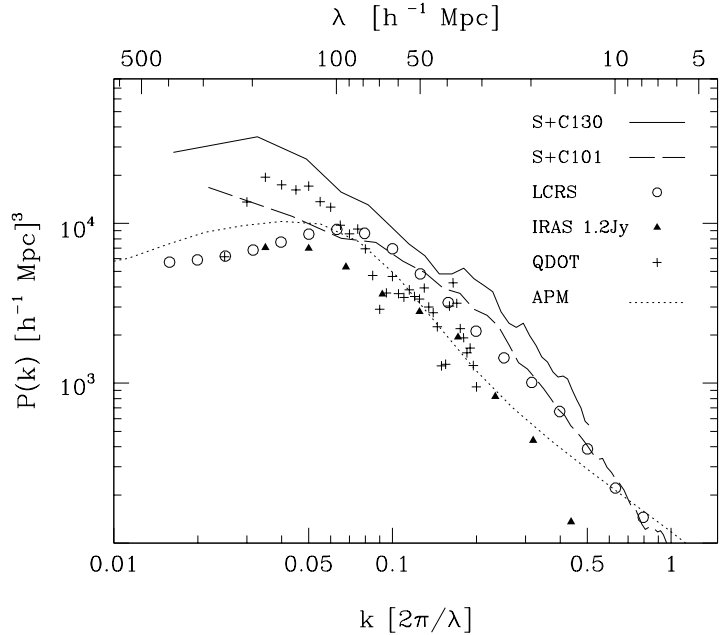

**Figure 1.** 3-D Power Spectra of Galaxy Surveys. Solid and dashed curves show redshift-space power spectra of volume-limited samples of the optically-selected SSRS2+CfA2 survey. Circles show the power spectrum of the Las Campañas Redshift Survey. Triangles and crosses show the redshift-space power spectra of flux-limited samples of the 1.2Jy and QDOT (1/6 of the 0.6Jy sample) IRAS surveys. The dotted line is the real-space power spectrum inferred from the angular correlations of the APM catalog.

## 2 Current Estimates of the 3-D Power Spectrum

Using standard estimation techniques (as described in section 3), the 3-D redshift-space power spectrum has been estimated for a variety of galaxy redshift samples. Figure 1 shows power spectra estimated from two volume-limited samples of the combined SSRS2+CfA2 survey [7], the Las Campañas survey [18], the IRAS 1.2Jy survey [10], the QDOT survey of IRAS galaxies [9], as well as the real-space power spectrum inverted from the angular correlation function of the APM catalog [1]. The shapes of the power spectra for the SSRS2+CfA2 samples, the Las Campañas survey, and the IRAS 1.2Jy sample are consistent within the errors. However the amplitudes of the power spectra for optical (SSRS2+CfA2 and Las Campañas) and infrared selected (IRAS 1.2Jy and QDOT) samples spectra differ quite significantly, as do the power spectra for SSRS2+CfA2 sample with different absolute-magnitude limits ($M < -20.3 + 5\log h$ and $M < -19.7 + 5\log h$ for the 130 and 101 samples, respectively), suggesting that the bias in clustering amplitude of different species extends over the full range of wavelength scales [20]. The QDOT survey exhibits a large excess in power on scales $\lambda > 75 h^{-1}$ Mpc over the shallower but more densely sampled 1.2Jy IRAS sample. This discrepancy may be caused by an extreme overdensity in the Hercules cluster, contained in the QDOT sample [27].

The APM curve in Figure 1 is an estimate of the *real-space* power spectrum, obtained by inverting the angular correlation function. The corresponding redshift-space spectrum would be steeper than this, and therefore in disagreement with the other optically-selected power spectra plotted here. Luminosity bias might explain this discrepancy: if the clustering amplitude is indeed an increasing function of galaxy luminosity, then the power spectrum inferred from angular correlations (which, by definition, use an apparent-magnitude limited sample) will have too steep a slope because the clustering amplitude increases with depth and therefore with wavelength scale.

These power spectrum measurements yield good constraints on models with CDM-like power spectra, but are insufficient to differentiate among the broad classes of contending models. The data can be well fit by a CDM power spectrum with $\Omega h \approx 0.25$ [17], [21]. The power spectrum of the "standard" $\Omega = 1, H_0 = 50 \text{kms}^{-1}$ model is excluded due to an excess of small vs. large-scale power. Due to the strong influence of peculiar velocities in this model, the shape of the redshift-space power spectrum is roughly correct, but the amplitude is too high (when normalized to COBE, this model requires $\sigma_8 = 1.4$ and, therefore, *anti-biasing* of all but the very brightest galaxies [13]). However, several alternative models predict power spectra with nearly the same shape and the correct normalization, among which the current data do not strongly discriminate. The list of candidates includes (but is not limited to) CDM models with non-zero cosmological constant, open universe CDM, mixed (cold plus hot) dark matter models [24], and warm plus hot dark matter [19].

To further constrain cosmological models, we must (1) close the gap between the scales probed by galaxy surveys and COBE, (2) measure the detailed shape of the galaxy power spectrum, (3) determine the dependence of clustering on galaxy species, and (4) quantify the anisotropy of clustering in redshift space caused by peculiar motions of galaxies. Comparison of power spectra for currently competing models (e.g., Fig. 1 of ref [26]) shows that the shapes of these spectra differ most strongly on scales near and beyond the peak of the spectrum. Unfortunately, the observed spectra plotted in Figure 1 all have resolution of $\delta k \sim 0.02$ or worse. Although there is some evidence of features in the observed power spectrum (e.g., the feature at $\lambda \sim 30 h^{-1}$ Mpc in the CfA2 and SSRS2 power spectra [7] and the peaks at $\lambda \sim 30 h^{-1}$ and $128 h^{-1}$ Mpc seen by [3]), the resolution of the power spectrum near the peak of the galaxy power spectrum is too poor to detect subtle features such as those produced in, e.g., models with a large baryonic mass component [15]. The turnover in the power spectrum seen in Figure 1 at $k \sim 0.03$ is similarly uncertain. More accurate probes of scales $\sim 100 h^{-1}$ Mpc and greater are necessary to confidently test for features in the power spectrum on scales where physical processes near the time of matter-radiation equality would leave their imprint and to compare galaxy clustering and the amplitude of mass clustering implied by CMB anisotropy measurements. The latter, along with knowledge of the dependence of clustering on galaxy selection, will elucidate the relationship between clustering of mass and light in the universe. Furthermore, measurement of the anisotropy of clustering in redshift space on the largest (and, therefore, presumably linear-growth) scales can yield a direct measurement of the mean cosmic density [16], [14], [5].

To increase the largest scales that we probe and the resolution of these measurements, we must survey a larger volume of the universe. Several ongoing and planned surveys promise to yield better constraints on the fluctuation spectrum, but the geometries of these samples pose a challenge to standard methods of power spectrum estimation. Deeper surveys of this type that are completed, or are soon to be, include pencil beam surveys [3], [4] and several deep slice surveys: the Las Campañas [25], Century [12], and ESP [29] surveys. Within the next two years we also expect results from the AAT 2df survey and the Sloan Digital Sky Survey (SDSS hereafter). Most of the sensitivity of the AAT survey to large-scale fluctuations will result from an ensemble of 100 randomly spaced pencil beams of 400 galaxies each. Over its five year duration, the SDSS will obtain redshifts for $10^6$ galaxies over a contiguous area of $\pi$ steradians in the North Galactic Cap, and therefore have a rather simple geometry. Figure 2 shows the uncertainty on the power spectrum that is expected for the full survey. However, the geometry of earlier partial data (e.g., $2 \times 10^5$ galaxies over some set of narrow stripes on the sky in the first year), and the survey in the South (three $2°\!.5 \times 100°$ stripes), will be more complex. Our goal is to extract the best possible estimate of the power spectrum from such partial data as the survey progresses.

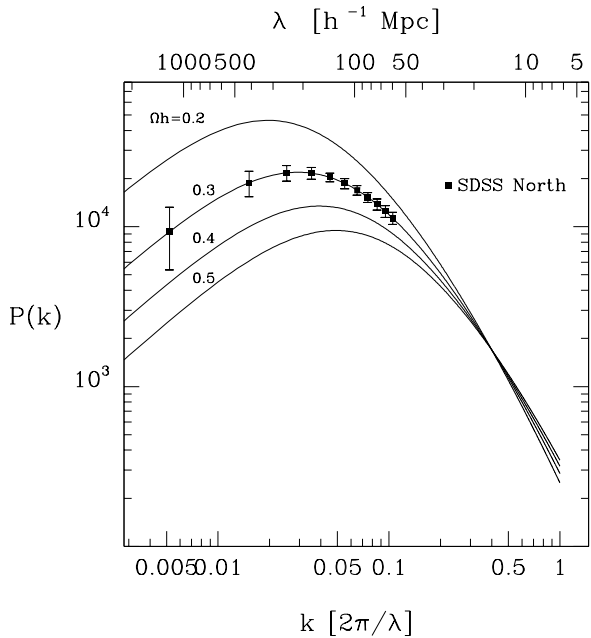

**Figure 2.** The $1\sigma$ uncertainty expected for a volume-limited (to $M^*$) sample of the SDSS northern redshift survey, assuming Gaussian fluctuations and a $\Omega h = 0.3$ CDM model, compared to power spectra for CDM with different $\Omega h$. Error bars on smaller scales are of similar or smaller size than the symbols.

## 3  Limitations of Standard Estimation Techniques

Standard methods for estimating the power spectrum all follow the same basic scheme. We directly sum the planewave contributions from each galaxy,

$$\tilde{\delta}(\mathbf{k}) = \frac{1}{\sum_j w(\mathbf{x}_j)} \sum_j w(\mathbf{x}_j) e^{i\mathbf{k}\cdot\mathbf{x}_j} - \tilde{W}(\mathbf{k}), \qquad (1)$$

where $w(\mathbf{x}_j)$ is the weight given to the $j^{th}$ galaxy and we subtract $\tilde{W}(\mathbf{k})$, which is the contribution to each mode from the finite survey window ($W(\mathbf{x}) = 1$ inside the survey and 0 elsewhere) and the selection function $\bar{n}(\mathbf{x})$. Next we compute the square of the modulus of each Fourier coefficient and subtract the power due to shot noise,

$$\hat{P}(\mathbf{k}) = \frac{|\tilde{\delta}(\mathbf{k})|^2 - \sum_j w^2(\mathbf{x}_j) / \left[\sum_j w(\mathbf{x}_j)\right]^2}{\frac{1}{(2\pi)^3} \int d^3 k' \, |\tilde{W}(\mathbf{k}')|^2}, \qquad (2)$$

and average these estimates over a shell in k-space to yield an estimate of $P(k)$. The denominator of equation (2) enforces the convention that $P(k)$ has the units of volume. Methods vary in the details (compare refs [20], [10], [9]), including the weights applied to each galaxy, how the window function of the survey is computed, corrections (or lack thereof) for the damping of large-scale power (analogous to the integral constraint on the correlation function – see below), and attempts to deconvolve the true power from the window function of the survey.

This method has several weaknesses that become even more serious when applied to surveys with complex geometry and sampling. A critical problem is that the basis functions of the Fourier expansion (plane waves) are not orthonormal over a finite non-periodic volume. Following equations (1) and (2), the power measured at any wavenumber is a convolution of the true power with the window function of the survey,

$$\hat{P}(\mathbf{k}) = \int d^3k' P(\mathbf{k}')|W(\mathbf{k}-\mathbf{k}')|^2. \qquad (3)$$

The estimates of power at different wavenumber have a covariance that depends on the shape of the survey volume. If the survey is oddly-shaped, then estimates $\hat{P}(\mathbf{k})$ with the same $k = |\mathbf{k}|$ but different direction $\hat{\mathbf{k}}$ sample different ranges of wavenumber because $W(\mathbf{k})$ is anisotropic. Averaging over a shell in $k$-space combines power estimates with varying bandpass and, therefore, different signal-to-noise ratios.

Further complications arise when we consider how to optimally weight the galaxies (as in eq. [1]) in different regions of a survey. The signal-to-noise for detection of clustering depends on the sampling rate of galaxies in the survey, which may vary due to survey strategy (e.g., the Las Campañas survey, which observes the same number of galaxies in each plug plate field), extinction (for a survey that includes galaxies to a fixed apparent magnitude if uncorrected for extinction), combining different surveys into a single sample, or simply because the selection function varies with distance (in the case where we analyze apparent-magnitude limited samples). In general, the set of weights applied to the galaxies should vary with the wavenumber $\mathbf{k}$ being probed. If the sampling density of galaxies varies with position on the sky, as in the Las Campañas survey, then the variation with wavenumber of the weight per galaxy yields a different pattern on the sky for each mode. Ignorance of this variation with wavenumber of the weighting scheme yields estimates of power with unnecessarily poor signal to noise.

Uncertainty in the mean density limits our ability to detect fluctuations on very large scales. In equation (1), when we subtract the contribution of the window function to each Fourier mode, we attempt to subtract the spike at $k=0$ in the true power, which is due to the non-vanishing mean density of galaxies. Because this spike is convolved with the window function of the survey and because we typically estimate the mean density from the sample itself (which forces $\langle \delta(\mathbf{x}) \rangle = 0$ within the survey), we erroneously subtract the product of the window function with the component of clustering signal on the scale of the survey, $|W(\mathbf{k})|^2 \langle |\delta_0|^2 \rangle$. A better method would be to simultaneously estimate both the mean density and the power spectrum.

Model testing is made difficult by the non-orthonormality of the Fourier modes, ambiguity about the optimal number of modes to include in the analysis, and necessary assumptions about the probability distribution of the measured power per mode. In principle, we could test models by computing the full covariance matrix of the power spectrum estimates for each model in consideration and compare their likelihoods, as approximated in ref [9]. This procedure requires that we repeatedly invert large, highly nondiagonal matrices. The size of the matrices could be reduced if we choose a limited set of modes, but this method does not specify the optimal set of power estimates; nearby modes have large covariance, but we lose statistical power and resolution if we sample too few. Finally, this method requires that we know the covariance matrix of the *power* per mode (which depends on fourth-order moments of the galaxy density) and the probability distribution of these fourth-order fluctuations for every model under consideration. To test the likelihood that the observations arise from a model with a particular power spectrum, we only require prediction of the covariance matrix of the expansion coefficients themselves, and knowledge of the probability distribution of second-order fluctuations in the density. A good choice of basis functions eases calculation of this matrix and increases the statistical power of the likelihood function.

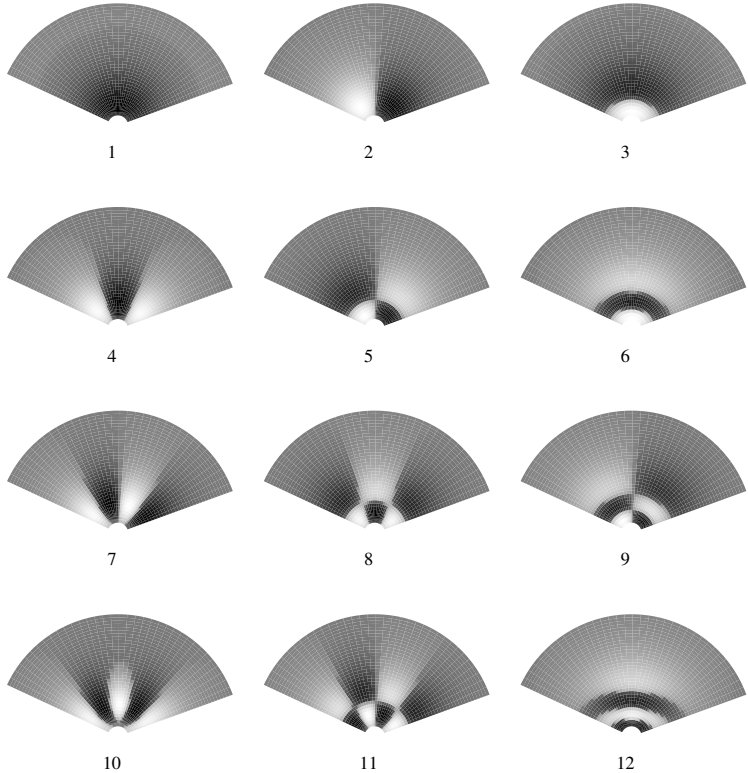

**Figure 3.** Eigenmodes of the apparent-magnitude limited CfA slice, formed by assuming the selection function and power spectrum measured for the CfA2 survey. This slice covers the region $29°\!.5 \leq \delta \leq 32°\!.5$, $8^h \leq \alpha \leq 17^h$, and we restrict the redshift range to $10h^{-1}\mathrm{Mpc} \leq r \leq 120h^{-1}\mathrm{Mpc}$. We plot the twelve modes with largest expected signal-to-noise ratio. These functions closely resemble the multipole moments of the density field, and are most sensitive to structure near the peak of the redshift distribution $r \sim 55h^{-1}\mathrm{Mpc}$.

## 4  The Karhunen-Loève Transform

Rather than make small modifications to the standard method for power spectrum estimation, we begin anew and derive the complete set of spatial functions that optimally weight the observed data in order to estimate second-order clustering properties of the galaxy distribution, and describe how this expansion naturally leads to straightforward likelihood analysis of proposed models [31]. The basis functions that we choose are the eigenvectors of the correlation matrix of galaxy density fluctuations. Expansion of the observed density field in this basis is known as the Karhunen-Loève transform (K-L hereafter; see, e.g., refs [28], [23] for discussion of the discrete and continuous forms of this transform, respectively). See ref [2] for a similar analysis of the COBE DMR maps.

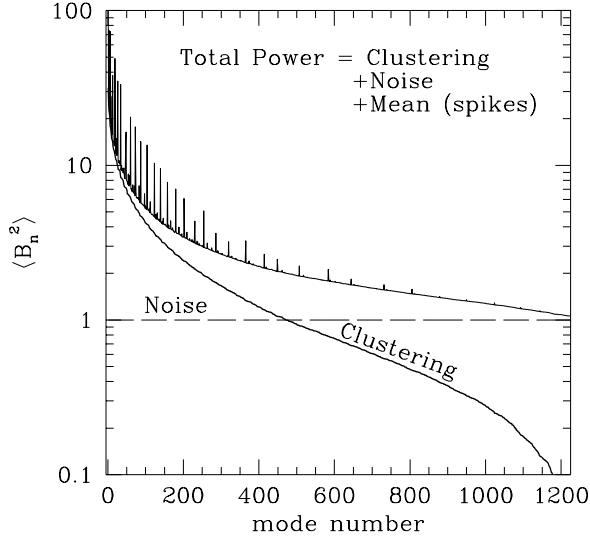

**Figure 4.** Expectation value of the power per mode for the K-L expansion of the CfA slice, where the modes are ordered by decreasing signal-to-noise ratio. The total power (upper solid line) is the sum of the clustering signal (lower solid line), noise (long-dashed line), and the mean density (spikes in the total power curve).

Dividing the survey volume into $M$ cells with volume $V_i$, we observe galaxy counts $d_i$, which form a vector $\mathbf{D}$. The correlation matrix of the count fluctuations $\mathbf{D} - \langle \mathbf{D} \rangle$ has elements

$$
\begin{aligned}
R_{ij} &= \langle (d_i - \langle d_i \rangle)(d_j - \langle d_j \rangle) \rangle \\
&= n_i n_j \xi_{ij} + \delta_{ij} n_i + \epsilon_{ij}
\end{aligned}
\quad (4)
$$

where $n_i \equiv \langle d_i \rangle$, $\delta_{ij} = 0$ for $i \neq j$, $\epsilon_{ij}$ is the correlation matrix for other sources of noise, and

$$\xi_{ij} \equiv \frac{1}{V_i V_j} \int d^3 x_i \int d^3 x_j \, \xi(\mathbf{x_i}, \mathbf{x_j}). \quad (5)$$

The three terms in equation (4) are the contributions from clustering of galaxies, shot noise, and extra variance due to, e.g., magnitude errors or uncertainty in the luminosity function. We assume a model for $\xi$ that is consistent with previous observations. The correlation function includes the redshift space distortions, $\xi(\mathbf{x}_i, \mathbf{x}_j) = \xi(r_p, \pi, R)$, where $r_p$ is the projected separation, $\pi$ the line of sight separation, and R the distance of the pair of cells from the observer.

The division of space into cells is rather arbitrary, therefore we must ensure that our derivation of the eigenmodes does not sensitively depend on the choice of pixellation. To derive the eigenmodes, we first apply a whitening transformation to the binned galaxy counts, $\mathbf{D}' = \mathbf{N}^{-1/2} \mathbf{D}$, where the $N_{ij}^{-1/2}$ are the square roots of the inverse of the noise component of the correlation matrix ($\mathbf{R} = \mathbf{S} + \mathbf{N}$, where $\mathbf{S}$ and $\mathbf{N}$ are the signal and noise correlation matrices, respectively). For the case of shot noise only ($\epsilon_{ij} = 0$ in eq. [4]), this yields a whitened correlation matrix,

$$\mathbf{R}' = \mathbf{N}^{-1/2} \mathbf{R} \mathbf{N}^{-1/2}, \quad (6)$$

with elements

$$R'_{ij} = n_i^{-1/2} [n_i n_j \xi_{ij} + \delta_{ij} n_i] n_j^{-1/2}$$

$$= n_i^{1/2} n_j^{1/2} \xi_{ij} + \delta_{ij}. \tag{7}$$

The K-L transform of the observed galaxy counts has coefficients

$$B_n = \mathbf{\Psi}_n^\dagger \mathbf{N}^{-1/2} \mathbf{D}, \tag{8}$$

where the $\mathbf{\Psi}_n$ solve the eigenvalue problem $\mathbf{R}'\mathbf{\Psi}_n = \lambda_n \mathbf{\Psi}_n$. The signal and noise are uncorrelated in this basis and the covariance matrix of these coefficients is diagonal. Figure 3 shows the first 12 eigenmodes for the geometry, selection function, and power spectrum of the first CfA slice [8].

Figure 4 shows the expected power per mode of the K-L transform, analogous to the power spectrum of the Fourier expansion. These modes are sorted by eigenvalue, which (because the expectation value of the noise power is unity for every mode) is equivalent to sorting by signal-to-noise ratio. The total power per mode is

$$\begin{aligned}
\langle B_n^2 \rangle &= \langle (\mathbf{\Psi}_n^\dagger \mathbf{N}^{-1/2} \mathbf{D})(\mathbf{\Psi}_n^\dagger \mathbf{N}^{-1/2} \mathbf{D})^\dagger \rangle \\
&= \mathbf{\Psi}_n^\dagger \langle \mathbf{D}' \mathbf{D}'^T \rangle \mathbf{\Psi}_n \\
&= \mathbf{\Psi}_n^\dagger \mathbf{R}' \mathbf{\Psi}_n + \mathbf{\Psi}_n^\dagger \langle \mathbf{D}' \rangle \langle \mathbf{D}' \rangle \mathbf{\Psi}_n \\
&= \mathbf{\Psi}_n^\dagger (\mathbf{S}' + \mathbf{N}' + \mathbf{E}') \mathbf{\Psi}_n.
\end{aligned} \tag{9}$$

The contributions from the mean density, noise, and clustering signal are

$$\begin{aligned}
E_n^2 &= \mathbf{\Psi}_n^\dagger \mathbf{E}' \mathbf{\Psi}_n = \langle B_n \rangle^2 \\
&= \left( \sum_{i=1}^M \Psi_n(\mathbf{x}_i) n_i^{1/2} \right)^2 \\
&= \left( \int d^3 x \, F_n(\mathbf{x}) \bar{n}^{1/2}(\mathbf{x}) \right)^2, 
\end{aligned} \tag{10}$$

$$N_n^2 = \mathbf{\Psi}_n^\dagger \mathbf{I} \mathbf{\Psi}_n = 1, \tag{11}$$

$$\begin{aligned}
S_n^2 &= \mathbf{\Psi}_n^\dagger \mathbf{S}' \mathbf{\Psi}_n \\
&= \sum_{i=1}^M \sum_{j=1}^M \Psi_n^*(\mathbf{x}_i) \Psi_n(\mathbf{x}_j) n_i^{1/2} n_j^{1/2} \xi_{ij} \\
&= \int \frac{d^3 k}{(2\pi)^3} |\tilde{G}_n(\mathbf{k})|^2 P(\mathbf{k}),
\end{aligned} \tag{12}$$

where we take the limit $M \to \infty$ and $V_i \to 0$ and define the discrete approximations to the continuous functions of position $F_n(\mathbf{x}) = \mathbf{\Psi}_n(\mathbf{x}_i)/V_i^{1/2}$ and $G_n(\mathbf{x}) = \mathbf{\Psi}_n(\mathbf{x}_i) \bar{n}^{1/2}(\mathbf{x})/V_i^{1/2}$ for $\mathbf{x} \in V_i$. Each mode samples the power spectrum with the window function $|\tilde{G}_n(\mathbf{k})|^2$. As we increase the volume covered by a survey, the Fourier windows of its eigenmodes narrow and we probe the fluctuations with increasing resolution. In the limit of an infinite volume, these windows approach delta functions, and the K-L eigenmodes become identical to the plane waves of the Fourier expansion, modulated by weighting to account for the variation with distance of the selection function.

We test proposed models for galaxy clustering in a Bayesian fashion, which includes evaluation of the likelihood of the data under a particular hypothesis. Restricting the analysis to large-scale modes, the likelihood is approximately

$$\mathcal{L}(\mathbf{B} \mid \text{model}) = (2\pi)^{-M/2} |\det \mathbf{C}_{\text{model}}|^{-1/2} \exp\{-(\mathbf{B} - \langle \mathbf{B} \rangle_{\text{model}}) \mathbf{C}_{\text{model}}^{-1} (\mathbf{B} - \langle \mathbf{B} \rangle_{\text{model}})^{\text{T}}/2\}, \tag{13}$$

where $M$ is the number of eigenmodes, and $\langle\mathbf{B}\rangle_{\text{model}}$ and $\mathbf{C}_{\text{model}}$ are predicted by the clustering model. In ref [31], we derive the K-L transform from the requirement that the confidence region of power spectrum parameters be as small as possible, therefore this basis set is *optimal* for testing clustering models. Note that we can explicitly vary the estimated mean density, rather than subtract the mean *a priori*, and thereby avoid the large-scale power damping described above. This method requires an initial guess at the power spectrum (to construct the eigenmodes), but the form of the eigenmodes does not depend sensitively on this assumption, and we can easily iterate the process.

The K-L eigenmodes form a complete basis for representing the observations, with no loss of phase information, and thus may be useful for analyses other than power spectrum estimation. One such application is smoothing of the density field by removal or suppression of modes that sample short wavelength scales. This form of smoothing could be used for studies of the morphology and topology of large-scale structure and for identification of superclusters in sparse data. Another application is optimal reconstruction of the galaxy density field, facilitated by the signal-to-noise properties of the eigenmodes (cf. refs [2] and [11] regarding Wiener filtering for reconstruction of the CMB anisotropy and the galaxy density field, respectively). When applied to spectroscopic observations of galaxies, the K-L transform yields an elegant means of spectral classification [6].

## 5 Conclusions

Statistical measures of the large-scale structure revealed by redshift surveys of the nearby universe and measurement of the CMB anisotropy successfully narrow the range of acceptable theoretical models, ruling out, for example, the previously 'standard' CDM model, and suggest consideration of several new models. The discriminatory power of galaxy clustering statistics and the increasing predictive power of theoretical models prompt planning for deeper complete redshift surveys such as the SDSS and the AAT 2df survey and the development of more sophisticated methods of analysis, such as the K-L transform.

We plan to apply this method for power spectrum estimation to a variety of data sets, including the pencil beam redshift surveys, early redshift observations from the Sloan Digital Sky Survey, and the spatial distribution of quasar absorption line systems. For future surveys (including planning for the SDSS), we can design the optimal geometry and sampling for the available survey resources, using the K-L transform as a method for estimation of the uncertainty in $P(k)$ for an arbitrary survey. We hope to use this transform method to combine all of the available galaxy redshift data and thereby obtain the best possible measurement of the power spectrum of galaxy density fluctuations.

**Acknowledgements.** I thank the many individuals who contributed to the work summarized above: Alex Szalay, for development of improved methods for power spectrum estimation; Luiz da Costa, Margaret Geller, John Huchra, and Changbom Park for analyses of the SSRS2 and CfA2 surveys. This research was supported in part by NSF grant AST-9020380 and the Seaver Foundation.